\documentstyle[prb,aps,multicol,epsfig]{revtex}

\begin{document}
\draft

\title{Mesoscopic phase separation in La$_2$CuO$_{4.02}$ - a $^{139}$La NQR
study}

\author{E.G. Nikolaev\cite{EN} and H.B. Brom}

\address{Kamerlingh Onnes Laboratory, Leiden University, P.O.Box 9504,
2300 RA Leiden, The Netherlands}

\author{A.A. Zakharov}

\address{RSC "Kurchatov Institute", Kurchatov sqr. 1, Moscow, 123182,
Russia}

\date{\today}

\maketitle

\begin{abstract}
In crystals of $\rm La_2CuO_{4.02}$ oxygen diffusion can be limited to such
small length scales, that the resulting phase separation is invisible for
neutrons. Decomposition of the $\rm ^{139}La$ NQR spectra shows the existence
of three different regions, of which one orders antiferromagnetically below
17~K concomitantly with the onset of a weak superconductivity in the crystal.
These regions are compared to the macroscopic phases seen previously in the
title compound and the cluster-glass and striped phases reported for the
underdoped Sr-doped cuprates.
\end{abstract}


\begin{multicols}{2}
\settowidth{\columnwidth}{aaaaaaaaaaaaaaaaaaaaaaaaaaaaaaaaaaaaaaaaaaaaaaaaa}

Inhomogeneous carrier and spin distributions in high-$T_c$
compounds might be related to phase separation and
stripes,\cite{Emery93,Zaanen98} and are intensively studied especially in
hole doped ${\rm La_2CuO_4}$. In underdoped $\rm La_{2-x}Sr_xCuO_4$ for
$x>0.06$ doping leads to spin-density wave order or stripe formation, as
seen by neutron scattering at various $x$-values.
\cite{Tranquada95,Suzuki98,Kimura99,Wakimoto99}  Recent NQR and NMR
studies have revealed new interesting features, like line intensity
suppression caused by spin/charge fluctuations, stripe condensation at low
temperatures under favorable pinning conditions, and the presence of
inequivalent copper sites attributed to a stripe
formation.\cite{Hunt99,Abu99,Curro99,Teitelbaum99} At very low $\rm Sr$
content ($x<0.02$) the magnetic properties are explained in terms of hole
segregation,\cite{Cho93,Chou93,Borsa95,Niedermayer98} and the formation of
a cluster spin glass (x=0.06).\cite{Julien99}

For ${\rm La_2CuO_{4+x}}$ the presence of the mobile oxygen dopants leads
to a macroscopic structural phase separation in antiferromagnetic (AF) and
superconducting (SC) regions in the concentration range $0.01<x<0.06$ (the
so called miscibility gap).\cite{Jorgensen88,Reyes93,Chou96} The oxygen
mobility is linked to lattice imperfections, e.g. it is strongly increased
by the presence of planar defects.\cite{Arrouy96} In ${\rm La_2CuO_{4+x}}$
single crystals prepared by the molten solution method \cite{Zakharov94}
oxygen mobility is very low due to the small number of defects and hence
the scale of the structural separation can be minimized. The single
crystal with $x=0.02$ prepared by this way appears homogeneous below 200~K
in X-ray and neutron studies,\cite{Pomjakushin98} although the composition
is inside the miscibility gap. In the crystal a superconducting transition
is observed around 15--17~K with a very weak diamagnetic
signal.\cite{Pomjakushin98,Nikonov00} According to
$\mu$SR\cite{Pomjakushin98} part of the sample becomes magnetically
ordered below the same temperature of 15~K, but neutron diffraction does
not see any long range magnetic structure. Apparently phase separation
happens, but the scale is too small to be visible by standard structural
methods.\cite{Pomjakushin98}

By performing NQR on the same ${\rm La_2CuO_{4.02}}$ single crystal,
we will demonstrate the existence of
three regions with different oxygen concentrations and show only one
(oxygen-poor) phase to have an antiferromagnetic transition. We evaluate
the size of the antiferromagnetic regions from the reduction of staggered
magnetization. By comparison with the macroscopic phases seen previously
in the title compound\cite{Reyes93} and the cluster-glass and striped
phases seen in $\rm La_{1.94}Sr_{0.06}CuO_{4}$ and La$_{2-x-
y}$Sr$_x$Eu$_y$CuO$_4$ we will discuss the influence of the character of
the local magnetic order on the NQR spectra.

\begin{figure}[h,t,b]
\begin{center}
\leavevmode
\epsfig{figure=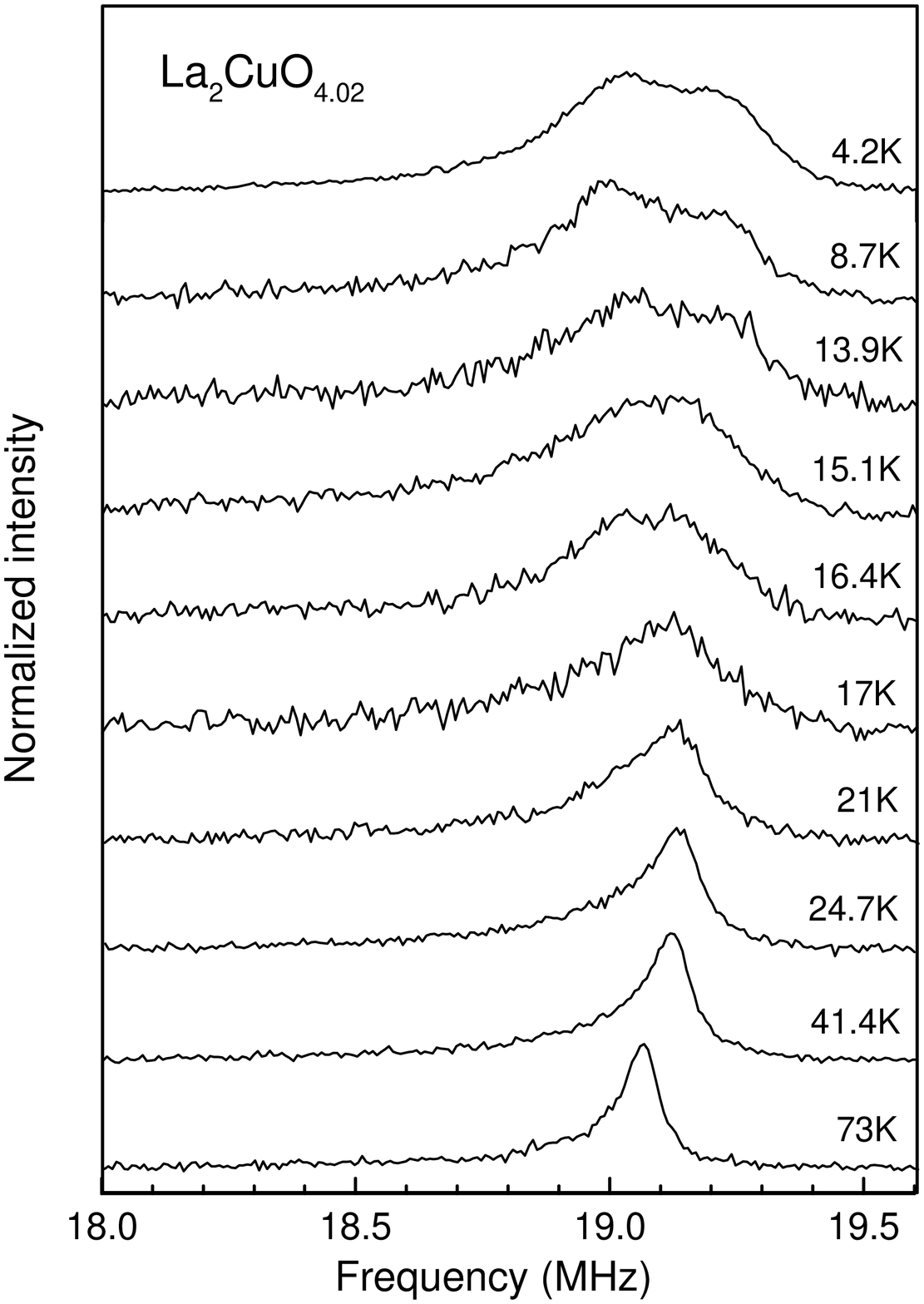,width=6.5cm,angle=0}
\end{center}
\caption{Normalized $^{139}$La NQR spectra for $m=7/2$ at various
$T$'s in the ${\rm La_2CuO_{4.02}}$ single crystal.}
\label{f1}
\end{figure}

\begin{figure}[h,t,b]
\centering\leavevmode \epsfig{figure=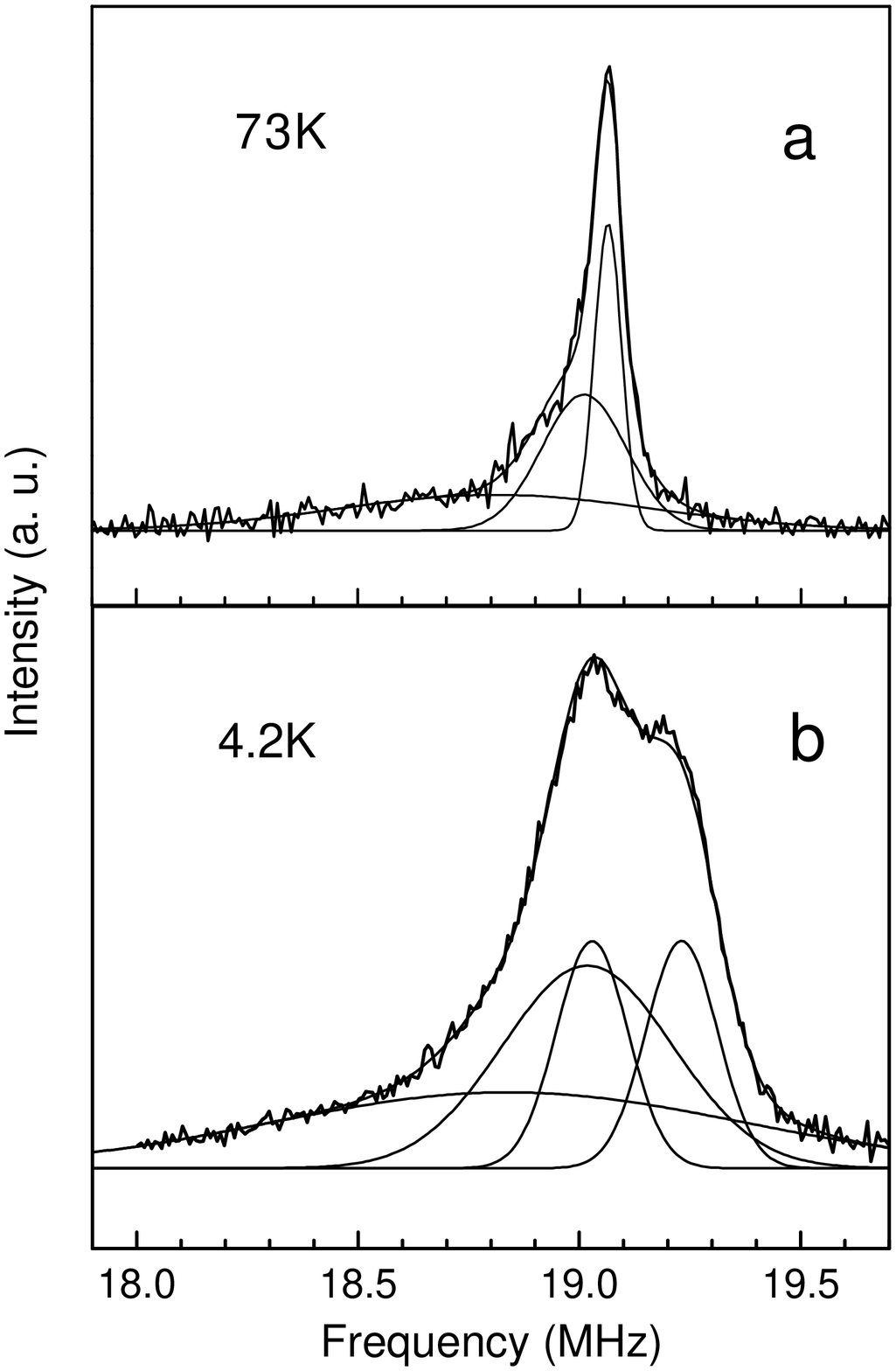,width=6.5cm,angle=0}
\caption{$^{139}$La spectral decomposition at 73~K and 4.2~K. Due to the
antiferromagnetic ordering the narrow line starts to split into two below
17~K.} \label{f2}
\end{figure}

\begin{figure}[h,t,b]
\centering\leavevmode \epsfig{figure=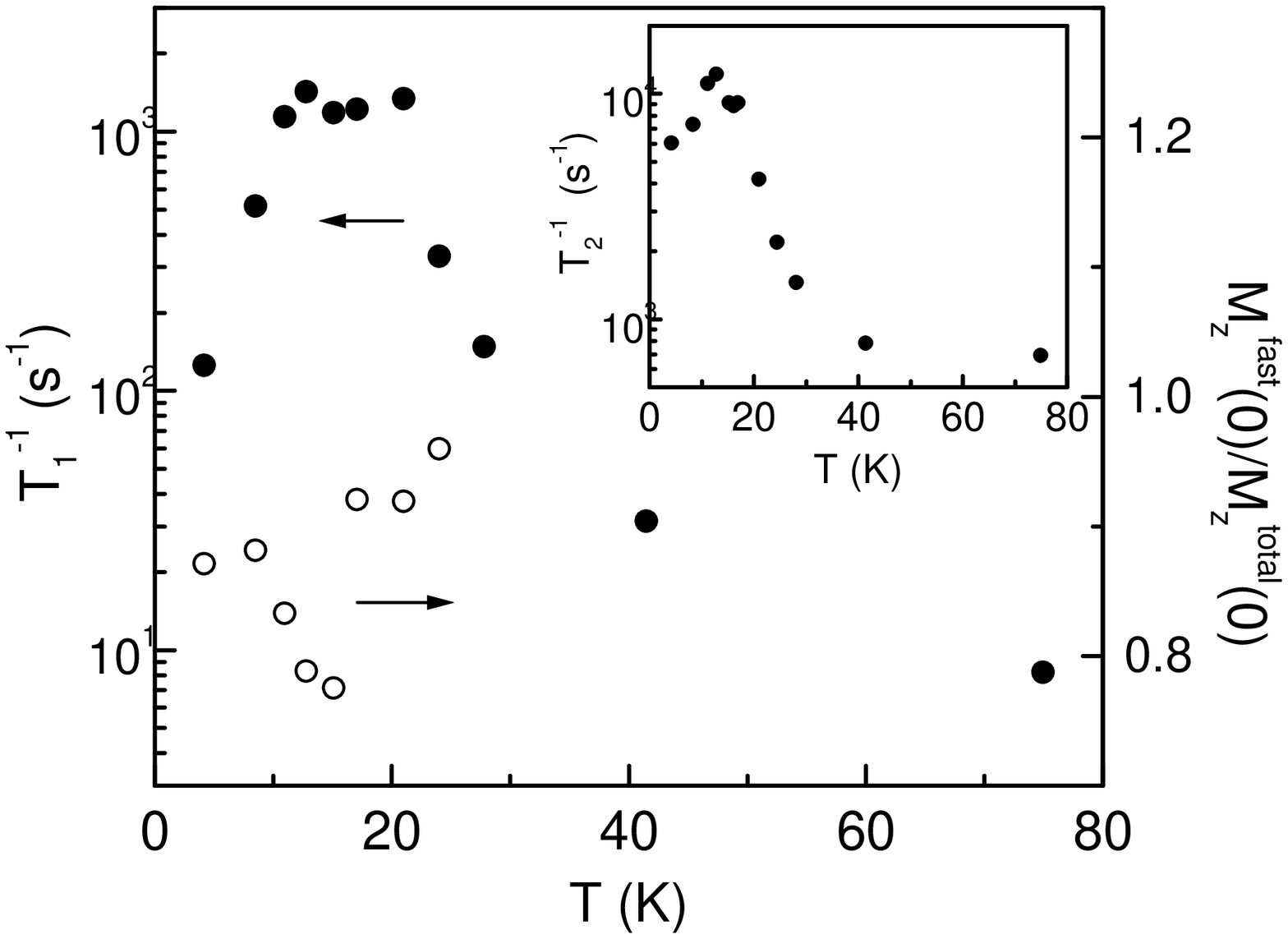,width=7.5cm,angle=0}
\caption{$^{139}$La spin-lattice relaxation rate vs. $T$. Closed circles
-- fast contribution, open circles -- relative value of the fast
contribution (scale on the right). $T_1$ (slow) is of the order of 0.1 s.
Inset: $^{139}$La spin-spin dephasing rate vs.  $T$.} \label{f4}
\end{figure}

The ${\rm ^{139}La}$ NQR spectra (Figs.\ref{f1},\ref{f2}) were taken for
the $\pm7/2-\pm5/2$ (referred to as $m=7/2$) transition by sweeping the
frequency. The nuclear spin-lattice relaxation rate $T_1^{-1}$
(Fig.\ref{f4}) was measured by monitoring the recovery of magnetization
after a single $\pi$ pulse. For the nuclear spin-spin relaxation (inset of
Fig.\ref{f4}) the standard spin-echo decay method was applied.
Fig.\ref{f1} shows the evolution of ${\rm ^{139}La}$ spectra with
temperature in the ${\rm La_2CuO_{4.02}}$ single crystal. A lower S/N
ratio around 15~K is due to the fast relaxation and partial wipe-out of
the signal in this region (see below). Differences in S/N in some adjacent
spectra are due to a different number of averages. The central part of the
spectrum broadens below 30~K. The splitting of the central line (see also
Fig.\ref{f2}b) below 17~K is a manifestation of magnetic ordering of the
Cu moments generating an internal magnetic field at the La site. Above the
magnetic transition the spectra are well described by the sum of three
Gaussian lines (labeled 1, 2 and 3 in sequence of increasing width). Below
17~K the spectra show the same structure but with a splitting of the
single narrow line 1 into two lines of equal intensity (1a and 1b) as a
result of the AF ordering. Figs.\ref{f2}a,b illustrate the results of a
spectral decomposition at 73~K and 4.2~K.

As already mentioned above, previous experiments showed the possibility of
the phase separation.\cite{Pomjakushin98} Our NQR spectra are in agreement
with this assumption. In addition our data show that separation is present
at temperatures much higher than that of the AF transition. We assign the
narrow line 1 (Fig.\ref{f2}a) to oxygen-poor (OP) regions, i.e. a phase
with low hole doping.  This phase is antiferromagnetically ordered below
17~K (two narrow lines in Fig.\ref{f2}b). The broad line (line 3) has to
be associated with oxygen-rich (OR) regions because of its absence in
samples with $x=0$. The large width (1~MHz) of this line is due to the
distortion of the electric field gradient at the La site because of the
presence of holes in the ${\rm CuO_2}$ planes leading to the distribution
of the oxygen octahedral tilts and the bond
lengths.\cite{Reyes93,Hammel93} The OR regions are likely responsible for
the superconductivity. The hole concentration in the CuO$_2$ planes does
not depend on the method of doping, which allows a comparison to the line
position data for La in La$_{2-x}$Sr$_x$CuO$_4$\cite{Kukovitsky95} (to see
the similarity with the copper data in ref.[\onlinecite{Teitelbaum99}],
one has to realize that extra holes shift the Cu resonance up, while
lowering the La frequency\cite{Oshugi94}). In La$_{2-x}$Sr$_x$CuO$_4$ the
line positions at 4.2~K obey the relation $\nu=(19.15 -5x)$~MHz.  Using
this expression and a hole to excess oxygen ratio of 2:1, the line
positions at 4.2~K of 19.13 MHz (midpoint of lines 1a and 1b), 19.02 MHz
and 18.84 MHz give oxygen concentrations of 0, 0.01 and 0.035 for lines 1,
2 and 3 resp. The $x$ dependence of $T_c$ in
La$_{2-x}$SrCuO$_4$\cite{Niedermayer98} gives an independent way to
estimate the hole concentration in the OR-regions. In the phase diagram
superconductivity is suggested to start around 0.06. $T_c$ = 15~K gives a
hole concentration in the OR phase near 0.07 or excess oxygen
concentration of 0.035, in agreement with the value found above. In the
macroscopically phase separated sample, the oxygen concentration in the OR
phase is 0.06.\cite{Reyes93,Chou96} This concentration difference is not
surprising since due to the restricted oxygen mobility the local oxygen
concentrations will be far from their equilibrium values. According to the
above given evaluation, line 2 belongs to regions with rather low oxygen
content, although it shows no sign of antiferromagnetic ordering. This
line possibly arises from the interface between OP and OR regions and
might be compared to the relatively narrow and weak line observed in
macroscopically phase separated ${\rm La_2CuO_{4.035}}$ besides the broad
metallic and splitted narrow antiferromagnetic lines.\cite{Reyes93} In our
case the weight in the spectrum will be enhanced due to the small sizes of
the correlated regions.

\begin{figure}[h,t,b]
\centering \leavevmode \epsfig{figure=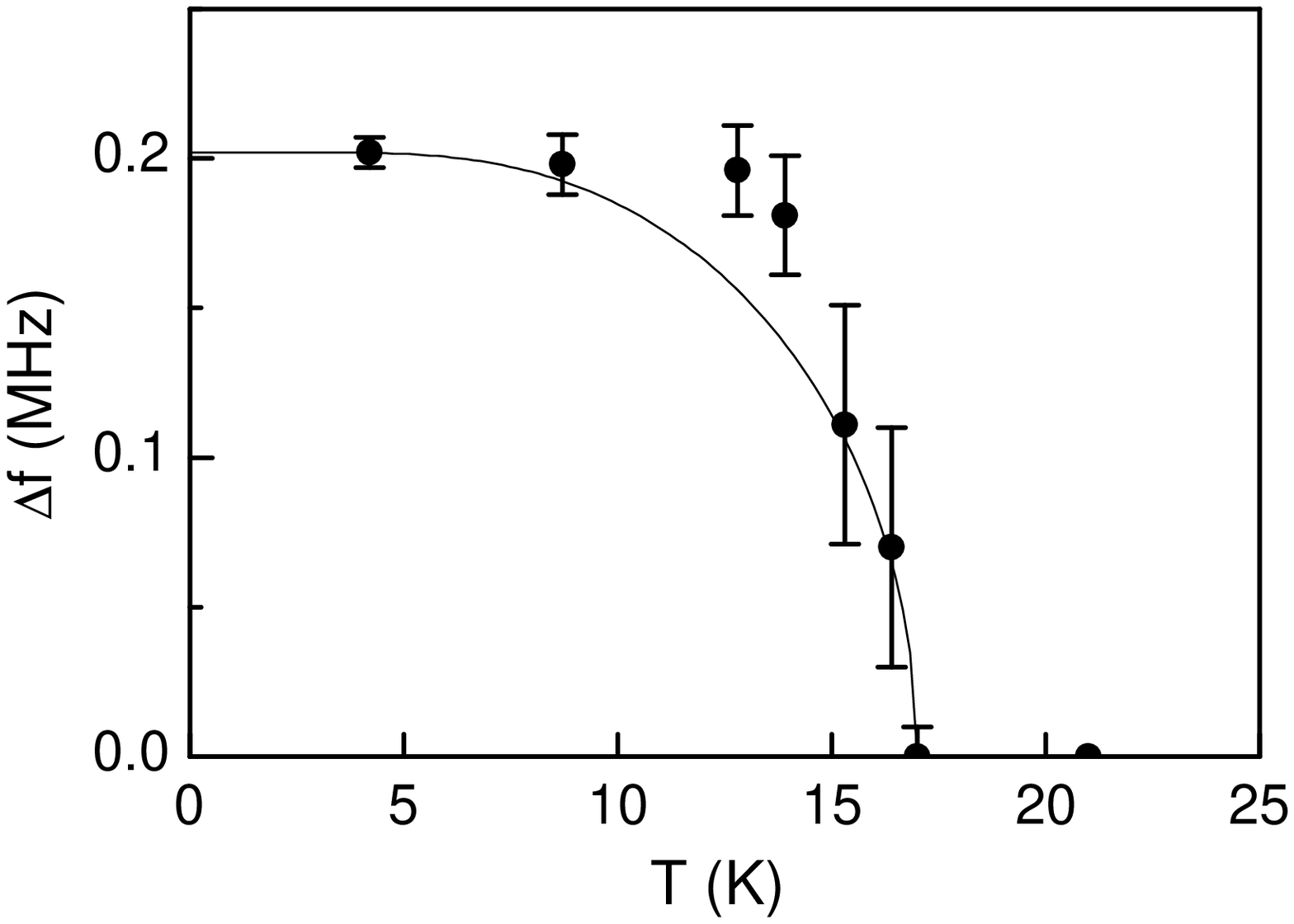,width=6.5cm,angle=0}
\caption{$^{139}$La NQR line 1 splitting for $m=7/2$ vs. $T$. Drawn line
is the mean field staggered magnetization for $S$=1/2 ($T_N$=17~K, $\Delta
f(0)$=202~kHz).} \label{f3}
\end{figure}

The splitting of the line 1 is almost $T$ independent from 14~K to 4.2~K
(see Fig.\ref{f3}) and just below the ordering temperature
$T_{AF}=17.0\pm0.5$~K seems to increase somewhat faster with lowering $T$
than the mean field behavior of the staggered magnetization for $S$=1/2
(drawn line on Fig.\ref{f3}). The line splitting (200~kHz) and hence the
internal field at the La-site at 4.2~K is 20\% lower than the 250~kHz
splitting in antiferromagnetic undoped ${\rm La_2CuO_4}$.
\cite{MacLaughlin94} This means a 20\% reduction of the staggered
magnetization and an ordered magnetic moment, that is 0.43~$\mu_B/{\rm
Cu}$ instead of $\langle\mu\rangle=$0.48~$\mu_B/{\rm Cu}$ in the undoped
compound. In the SDW--like AF ordered striped phase of the Eu and Sr doped
compound this value is $0.29\mu_B/{\rm Cu}$.\cite{Teitelbaum99} A
temperature dependent reduction of the staggered magnetization in the
system with low Sr doping has been explained by finite size
effects\cite{Fisher72} caused by microsegregation of the holes in domain
walls of flat domains.\cite{Cho93,Borsa95} At low temperatures this effect
vanishes due to the hole localization. In our case local field reduction
is observed even at 4.2 K and may be explained as a finite size effect
caused by a small grain dimension. At $T$ = 0~K the staggered
magnetization of a domain of size $L$ (infinite in the two other
directions) drops by 20\% for $L$ about 20 lattice spacings
(8~nm).\cite{Borsa95} For a quasi-two-dimensional domain (due to the large
anisotropy) the reduction should increase and hence this value has to be
considered as a lower limit. The neutron data\cite{Pomjakushin98} place an
upper limit of the same order of magnitude.  Thus an average grain size of
the order of 8~nm looks reasonable and confirms phase separation on a
mesoscopic rather than macroscopic scale.

The $T$ dependencies of the spin-lattice ($T_1^{-1}$) and spin-spin
($T_2^{- 1}$) relaxation rates are illustrated in Fig.\ref{f4} and its
inset. Both rates show a large peak at low temperatures.  Near the peak
temperature, the nuclear magnetization recoveries after the initial
magnetization reversal ($M_z(t)$) in case of $T_1^{-1}$, or the loss of
transversal magnetization ($M_x(t)$) in case of $T_2^{-1}$, are
characterized by stretched exponential time dependencies
$\exp[-(t/T_{1,2}]^{\alpha})$.\cite{relaxation} Very close to the maximum,
fits of $M_z(t)$ require a slow and fast time constant. For both $T_1$'s
$\alpha$ is about 0.5, while the quality of the $M_z(t)$-fit is found to
depend only weakly on the precise value for $T_1$ (slow) (of the order of
0.1~s). Closed circles in Fig.\ref{f4} show the temperature dependence of
the fast rate.  Below 17~K there is a sharp decrease of this contribution
(open circles in Fig.\ref{f4}).

\begin{figure}
\centering\leavevmode \epsfig{figure=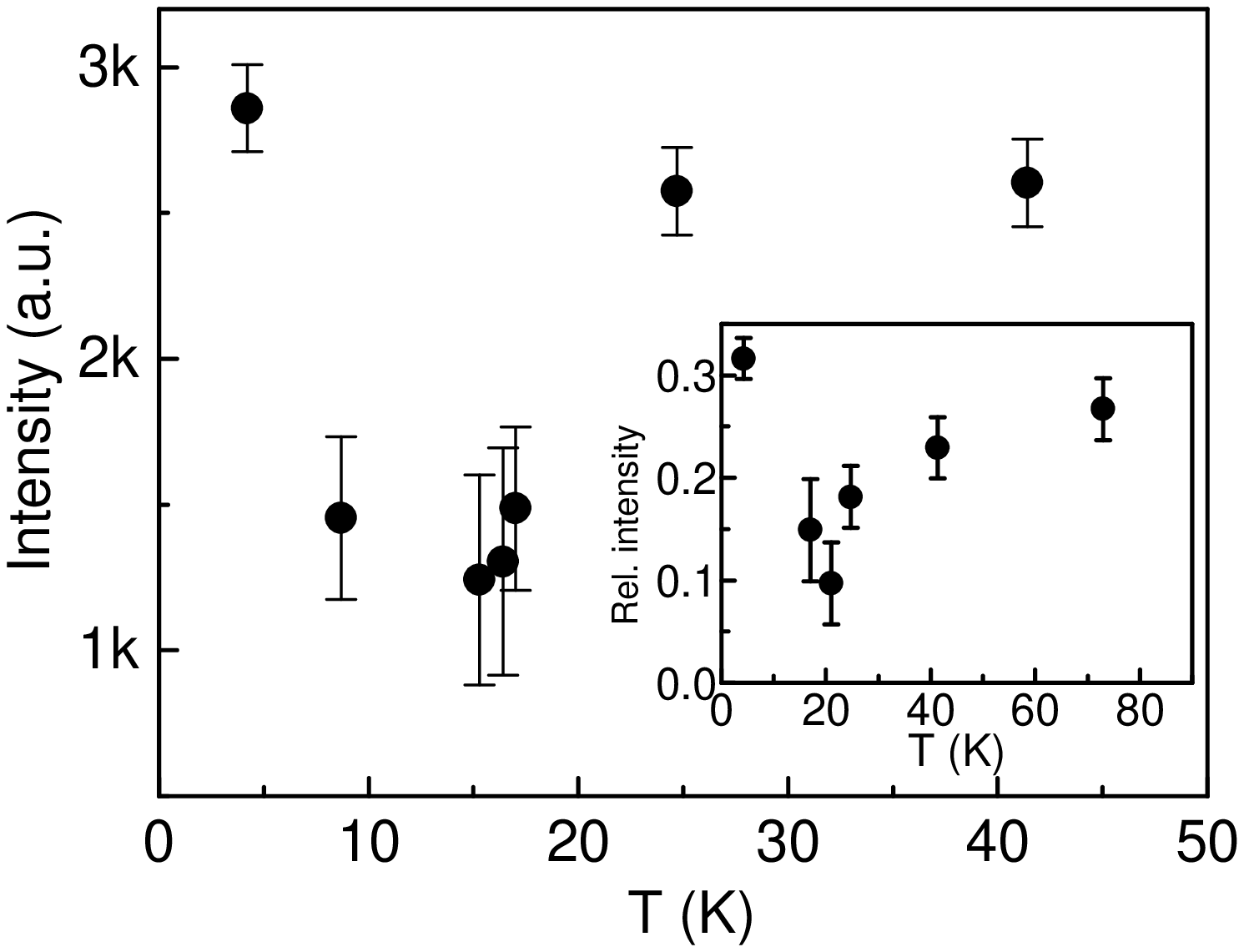,width=6.5cm,angle=0}
\caption{Corrected intensity of the ${\rm ^{139}La}$  spectrum vs. $T$.
Inset: Relative intensity of the narrow line vs. $T$.} \label{f5}
\end{figure}

The strong growth of the ${\rm ^{139}La}$ relaxation rates around 15~K is
explained as a result of hole localization and slowing down of hole
mediated spin fluctuations.\cite{Curro99,Chou93} Here in the same
temperature region the AF and SC transitions occur. The coincidence of
these phenomena likely leads to even more complicated behavior of nuclear
relaxation rates, e.g. the above mentioned decrease of the fast
contribution to the longitudinal relaxation just below 17K (Fig.\ref{f4}).
The occurrence of slow and fast contributions to $M_z(t)$ is typical for a
multiphase material. Further analysis is not pursued, as its precise
character might not only depend on the AF transition alone, but on the
coexisting SC transition as well. The other feature caused by the low
temperature relaxation behavior is a partial wipe-out of the La NQR signal
in the same temperature region. The total intensity multiplied by $T$ and
corrected for the echo decay factor (Fig.\ref{f5}) shows a decrease around
15~K by about a factor 3 indicating part of the nuclei to have dephased
before the echo signal could be measured.\cite{Curro99} The relative
decrease of the intensity of the narrow line (1) starts below 70~K (inset
of Fig.\ref{f5}) and becomes very pronounced below 30~K. It demonstrates a
difference in relaxation behavior in the OP and OR phases. The imprint on
other NQR features, if any, might be limited to the changes in the
stretched exponential recoveries (see above).

How do these results compare to those in the low doped Sr-cuprates?  In
$\rm La_{1.94}Sr_{0.06}CuO_{4}$\cite{Julien99} the La spectra (NMR and
NQR) show the existence of at least two sites below 200~K, of which one
site is ascribed to AF clusters. The change in the La NMR spectrum around
20 K seems to signal glass formation. Wipe-out effects are clearly seen
for Cu-NMR but no La-NQR wipe-out data are shown. In
La$_{1.48}$Sr$_{0.12}$Nd$_{0.4}$CuO$_4$ La wipe-out is found to be almost
complete.\cite{Teitelbaum00} In La$_{2-x-y}$Sr$_x$Eu$_y$CuO$_4$ with
$0.08<x<0.18$, Cu-NQR shows the presence of three inequivalent
sites,\cite{Teitelbaum99} which is also reminiscent to our case.  At low
doping at 1.3~K the Cu-intensity is strongly reduced compared to a $x=1/8$
sample due to wipe-out effects without affecting the relative strengths of
the three lines.  This comparison shows at least three major differences
of the NQR spectra: in La$_2$CuO$_{4.02}$ wipe-out effects on the La-site
are less severe, and mainly linked to the relaxation peak and the
antiferromagnetic transition below 17~K, line features are much sharper
than in the Sr-doped cuprates, and the line splitting (hence internal
field) in the AF state is almost the same as in the undoped case. The
large wipe-out values over very extended temperature regions of the
Sr-doped samples seem to be typical for mobile striped or glassy phases,
which are not expected in our compound.

The most intriguing feature is the coincidence of the AF and
superconducting transitions in the investigated single
crystal.\cite{Pomjakushin98} The AF ordering temperature is very low in
comparison with $T_N$'s of macroscopically phase separated ${\rm
La_2CuO_{4+x}}$ samples, which are usually higher than 200~K.\cite{Chou96}
The transition is sharp (Fig.\ref{f3}) and it looks as the AF state is
strongly depressed at high temperatures and is triggered by some reason at
low temperatures (the above mentioned relaxation behavior below 17K
supports this assumption). It has been argued that superconductivity
itself destabilizes the homogeneous metallic state and leads to the
formation of (super)conducting droplets weakly linked to each other and
separated by insulating (antiferromagnetic) regions.\cite{Gorbatsevich90}
Our system is already inhomogeneous above the AF and SC transitions
because the oxygen phase separation occurs at much higher temperatures but
the hole concentration in OR grains is far from optimal. Therefore the
occurrence of superconductivity in the OR grains might cause an additional
electronic redistribution, which favors the AF transition in the OP
grains. The other possibility is that both antiferromagnetism and
superconductivity are triggered by a common mechanism e.g. slowing down of
spin fluctuations coupled to hole motion occurring in the same temperature
region.

In summary in single crystal of La$_{2}$CuO$_{4.02}$ with limited oxygen
mobility the three different La sites seen by NQR are associated with
oxygen-poor, oxygen-rich and intermediate regions reminiscent the O-doped
macroscopically separated system and the Sr-doped glassy or striped
La$_{2}$CuO$_{4}$ compounds. Sharper lines, well defined transition
temperatures, internal fields close to bulk values and less severe
wipe-out effects are seen as the main difference of the NQR data with low
Sr-doped cuprates instead of O-doping. The evaluated OP grain size of the
order of 8 nm shows that the crystal is phase separated on a mesoscopic
scale. Although, like in the underdoped high-$T_c$ systems,
superconductivity itself is not directly reflected in the NQR data, some
features of the antiferromagnetic transition in the OP phase (a sharp
transition at low temperature with $T_N \sim T_c$ and $^{139}$La
relaxation peculiarities) are indicative for a possible coupling with the
superconducting transition in this system.

We gratefully acknowledge Issa Abu-Shiekah for his help in the experiments.
This work is supported in part by FOM-NWO, and NWO/INTAS-1010-CT93-0045.

\end{multicols}

\end{document}